\documentstyle[twocolumn,prl,aps,epsf]{revtex}
\begin{document}
\draft
\twocolumn[\hsize\textwidth\columnwidth\hsize\csname @twocolumnfalse\endcsname
\input{psfig}
\title
{Direct Observation of the Quantum Energy Gap in $S$=$\frac{1}{2}$
Tetragonal Cuprate Antiferromagnets}

\author{K. Katsumata$^a$, M. Hagiwara$^a$, Z. Honda$^a$\cite{honda}, 
J. Satooka$^a$, Amnon Aharony$^{b,c}$, 
R. J. Birgeneau$^{c,d}$, F. C. Chou$^c$, O. Entin-Wohlman$^b$, 
A. B. Harris$^e$, M. A. Kastner$^c$, Y. J. Kim$^c$,
and Y. S. Lee$^c$}

\address
{$^a$ RIKEN (The Institute of Physical and Chemical Research),
Wako, Saitama 351-0198, Japan}

\address
{$^b$ School of Physics and Astronomy, Sackler Faculty of
Exact Sciences, Tel Aviv University, Tel Aviv 69978, Israel}

\address
{$^c$ Center of Materials Science and Engineering, Massachusetts Institute of
Technology, Cambridge, Massachusetts 02139}

\address
{$^d$ University of Toronto, Toronto, ON, Canada, M5S 1A1}

\address
{$^e$ Department of Physics, University of Pennsylvania, Philadelphia,
Pennsylvania 19104}

\date{\today}

\maketitle

\begin{abstract}
Using an electron spin resonance spectrometer covering a wide range of
frequency and magnetic field, we have measured the low energy excitations of
the $S$=$\frac{1}{2}$ tetragonal antiferromagnets,
Sr$_{2}$CuO$_{2}$Cl$_{2}$ and Sr$_{2}$Cu$_{3}$O$_{4}$Cl$_{2}$.
Our observation of in-plane energy gaps
of order 0.1 meV at zero external magnetic
field 
are consistent with a spin wave calculation, which includes several
kinds of quantum 
fluctuations that remove frustration. Results agree with other experiments
and with
exchange anisotropy parameters determined from a five band
Hubbard model.
\end{abstract}

]

The fact that many systems containing copper oxide planes become high-$T_c$
superconductors when suitably doped \cite{BM86}
has led to a continuing effort to
understand in detail the magnetic properties of the undoped parent systems.
Most of these systems contain weakly coupled CuO$_2$ planes, and the 
$S=\frac{1}{2}$ spins  ${\bf S}_i$ on the Cu ions are well 
described as an antiferromagnet governed by the Hamiltonian
${\cal H} = \sum_{\langle ij \rangle} {\bf S}_i \cdot {\bf J}_{ij} 
\cdot {\bf S}_j$,
where 
${\bf J}_{ij}$ is the tensor for exchange interactions between ions
$i$ and $j$, and $\langle ij \rangle$ restricts the sum to pairs of
nearest neighboring Cu spins.  Above the N\'eel temperature $T_{\rm N}$, most
experiments can be fully explained by specializing to only intraplanar
isotropic coupling, so that ${\bf J}_{ij} = J{\cal I}$, where
${\cal I}$ is the unit tensor and $J \sim130$ meV\cite{rev}.
For isotropic coupling the spin-wave spectrum is
doubly degenerate and the spin-wave energy $\omega ({\bf q})$
goes to zero as $q \rightarrow 0$.  The existence of 3D long range
antiferromagnetic (AF) order, with 
$T_{\rm N} \sim 250-400$ K, 
and the existence of some non-zero spin-wave gaps at $q=0$
require an {\it anisotropic} exchange tensor in the plane, and/or
some 3D exchange coupling.
However, some of these small gaps, predicted by theory, have not yet
been observed experimentally.

Although some of the phenomena in the cuprates can be explained by a
classical treatment of anisotropic Heisenberg models which ignore
quantum fluctuations, the cuprates still have several interactions which
are {\it frustrated} at the mean-field level, and would lead to
a ground state degeneracy.
One example concerns the {\it in-plane exchange anisotropy} 
in the planar square
lattice common to most tetragonal cuprates. Symmetry implies that the
principal axes of ${\bf J}_{ij}$ in the plane are
parallel to the $i-j$
bond ($\parallel$), perpendicular to the bond in the plane
($\perp$), and perpendicular to the plane ($z$). Thus, the spin interaction
along this bond is of the form $J_\parallel S_{i\parallel}S_{j\parallel}
+J_\perp S_{i\perp}S_{j\perp}+J_zS_{iz}S_{jz}$. Indeed, a five-band
Hubbard model, in the limit when the on-site Coulomb repulsion dominates
the hopping matrix elements \cite{yildirim}, including both spin-orbit
and Coulomb exchange interactions, yields deviations of the principal
values $J_\alpha$ by a few parts in $10^4$ from their average, 
$J$.  The out-of-plane gap, $\omega_{\rm out}=5$ meV, observed in many
cuprates
\cite{rev}, is related to the out-of-plane anisotropy field,
$H_{\rm A}^{\rm out} \equiv (2J_\perp + 2J_\parallel-4J_z)S$:
\begin{equation}
h\nu \equiv \omega \equiv \sqrt{2H_{\rm E}H_{\rm A}} \equiv
\sqrt{8JS H_{\rm A}},
\label{gap}
\end{equation}
where $h$ is Planck's constant. 
(In what follows, we give $\nu$ in units of GHz, and $\omega$
in meV: 1 meV/$h=241.8$ GHz.)
Since $J_\perp \not= J_\parallel$, one
might expect a similar gap, $\omega_{\rm in}$,
for in-plane spin waves at ${\bf q}=0$.
However, for a classical spin
model at $T=0$, the sum over perpendicular bonds
yields an isotropic planar energy and hence $\omega_{\rm in}=0$.
This isotropy is removed by {\it quantum fluctuations},
and detailed calculations yield an in-plane anisotropy field
\begin{equation}
H_{\rm A}^{\rm in,K}= 8K/S=C (J_\parallel-J_\perp)^2/J, 
\label{HAin}
\end{equation}
where 
$C \approx 0.16$ 
and the quantum four-fold energy per unit cell
is ${\cal H}_4=-K\cos 4\theta$
($\theta$ is the angle between the staggered moment and a Cu--Cu bond, and
there are two Cu ions per cell)
\cite{yildirim}.  $H_{\rm A}^{\rm in,K}/S$ is
of order $1/S$ ($K \propto S$), emphasizing that  $H_{\rm A}^{\rm in,K}$
is a manifestation of quantum fluctuations \cite{renorm}.  The fact that
$H_{\rm A}^{\rm out}$ is linear in the exchange anisotropy, whereas
$H_{\rm A}^{\rm in,K}$ is quadratic in that small quantity, indicates
why the in-plane gap due to $H_{\rm A}^{\rm in,K}$
is too small to be detected by
inelastic neutron scattering, thus requiring the present electron
spin resonance (ESR) experiments.  

A second (and more familiar) example of frustration occurs in
materials like Sr$_2$CuCl$_2$O$_2$ (``2122"),
which has the body centered tetragonal K$_{2}$NiF$_{4}$ structure with the
CuO$_{2}$ layers in the $c$ plane\cite{VBG75};
each Cu couples to four equidistant Cu's in a neighboring plane.
For isotropic inter--plane exchange, the mean field sum of these four
interactions vanishes.  Nevertheless, below 
$T_{\rm N}\simeq$251 K, the spins have a well defined 
AF structure, with the easy axis believed to be parallel to
the [110] direction\cite{DV90}.  The magnetic structures of such
cuprates have been explained\cite{structure} by
considering, in addition to $K$,
two competing energies, which also relieve the frustration.
The first of
these is an effective {\it bilinear interplanar coupling} \cite{shender}
of the form $-j_{\rm eff} ({\bf S}_i \cdot {\bf S}_j)^2$ which is
generated by {\it quantum fluctuations} of the otherwise frustrated
isotropic
interactions.  This effective coupling favors colinearity of the spins
in neighboring planes. The second additional energy, $A_{\rm pdip}$, arises
from
the small interplanar ``pseudodipolar" exchange 
anisotropic interaction (not yet calculated), 
which  adds to the dipolar interaction $A_{\rm dip}$,
giving a contribution, 
\begin{equation}
H_{\rm A}^{\rm in,d} \equiv 4A/S \equiv
4(A_{\rm dip}+A_{\rm pdip})/S,
\label{Hd}
\end{equation}
to the in-plane anisotropy, where the parameter $A$ was defined
in Ref.  \onlinecite{structure}.

Sr$_2$Cu$_3$O$_4$Cl$_2$ (``2342") combines the above quantum effects.
In 2342, the CuO planes have an additional
Cu ion (denoted CuII) at the center of every alternate CuI plaquette.
The CuI subsystem shows AF
ordering at $T_{\rm NI}$$\simeq$380 K.  Although the isotropic molecular
field acting on the CuII sites from the CuI's vanishes (similar to the 
interplanar field in 2122),
the CuII subsystem exhibits a small ferromagnetic moment below
$T_{\rm NI}$, and shows AF ordering
at $T_{\rm NII}\simeq$40 K, with its staggered moment colinear to that
of the CuI's, i. e. along [110] (parallel to a CuI--CuI bond). 
Both the magnetic structure \cite{chou,kastner} and
spin-wave spectrum\cite{kim} have been explained by including all the
three mechanisms mentioned above.

Here we report on ESR measurements of the in--plane fluctuation
induced gap in both 2342 and 2122\cite{HO95}.
The  observation of these small gaps in the predicted range of energy gives
decisive confirmation both of the model, in which the gaps
are attributed (at least partly) to quantum fluctuations,
and also of our fundamental understanding of
the electronic structure of the cuprates which is used to
calculate $J_\alpha$.
Also, the magnetic field dependence of the gap
gives values for the anisotropic g tensor for the Cu spins, which
roughly agree with theory.

The single crystals of 2122 and 2342 used in this study were grown at MIT by
slow cooling from the melt.  The ESR measurements were performed using the
spectrometer installed in RIKEN\cite{MH96}.  Several Klystrons and Gunn
oscillators were used to cover the frequency range from 20 to 100 GHz.  We also
used a vector network analyzer, bought from the AB millimetre Company,
operating in the frequency
range 50-700 GHz.  Magnetic fields up to 20 T were generated with a
superconducting magnet from Oxford Instruments.  Since the ESR signals
in these samples were weak at low frequencies, we used resonant cavities at
respective frequencies below 70 GHz and a field modulation technique to
enhance the sensitivity.

Figure 1 shows typical ESR signals, observed at 70 K in a single
crystal of 2342 with the external magnetic field (${\bf H}$) 
parallel to the easy axis [110] at
the designated frequencies.  Because we use different resonant cavities for
different frequencies, a direct comparison of the absorption intensity is
difficult.  However, we see a tendency of the intensity (which is obtained
by a double integration of the spectrum with respect to $H$) to decrease with
decreasing frequency.  The intensity in the 2122 sample is much weaker than
that of the 2342 sample.

Since 2342 has two transitions, we measure the antiferromagnetic
resonance (AFMR) at two temperatures:
70 K, between $T_{\rm NI}$ and $T_{\rm NII}$, and 1.5 K ($<T_{\rm NII})$.
In Figs. 2 and 3
we plot the frequency ($\nu$) dependence of the resonance fields
observed in single crystals of 2342 and 2122, at the indicated temperatures
and field directions.
The experimental points constitute
separate branches, each showing clearly an energy gap at $H$=0. 
Each set of data has been fitted
by \cite{NYK55}
\begin{equation}
\nu (H)^2 = \nu (0)^{2} + (g \mu_B H/h)^2 ,
\label{nu}
\end{equation}
where $g$ is the g value for the corresponding orientation of ${\bf H}$, 
and $\mu_{\rm B}$ the Bohr magneton. 
The fitted coefficients, with their statistical errors, are
given in Table I.


Now we compare the experimental results with the theory, beginning with 2342
at 70 K.
Ignoring the small ferromagnetic moment (which only introduces a negligible
shift in the effective $K$), 
the intermediate phase ($T_{\rm NII}$$<$$T$$<$$T_{\rm NI}$)
is similar to the ``usual" AF phase of other cuprates. The AF ordering
of the CuI's generates only two low energy modes, $\omega_{\rm out}$
and $\omega_{\rm in}$,
given by Eq. (\ref{gap})
with $H_{\rm A}^{\rm out}$ and $H_{\rm A}^{\rm in,K}$.  The former has an
energy of order 5 meV, which 
is too large for our AFMR measurements (but
has been measured by neutrons in Ref. \onlinecite{kim}).
Using the experimental value of 
$\omega_{\rm in}$ at 70 K 
in Eqs. (\ref{gap}) and (\ref{HAin}) gives $K=\omega_{\rm in}^2/(64J)=
(5.2 \pm 0.3) \times 10^{-7}$ meV \cite{renorm}.
This roughly agrees with the static experimental value of Kastner
{\it et al.} \cite{kastner}, $K=(10 \pm 3) \times 10^{-7}$ meV.
Returning to Eq. (\ref{HAin}), these two values of $K$
imply that
$\delta J \equiv |J_\parallel - J_\perp|  = \sqrt{ 8JK/(SC)} 
\approx 0.08-0.11$ meV,
where we use $J$=130 meV\cite{kim} and $S=1/2$ \cite{renorm}. 
This value of $\delta J$ is a factor of two larger than that estimated
theoretically by Entin-Wohlman {\it et al.} \cite{entin}, for the geometry
of 2122. A larger value of $\delta J$ for 2342, compared to 2122,
could result from the difference in environment (due to insertion
of CuII's), from the uncertainties in the Hubbard model parameters
used in Refs.  \onlinecite{yildirim,entin},
or from higher order renormalizations\cite{renorm}.
Unlike the other gaps discussed below, $\omega_{\rm in}$ at 70 K for 2342
is purely due to fluctuations, which are of quantum origin at $T=0$; 
$H_{\rm A}^{\rm in,d}=0$ for the 
CuI spins (which sit on top of each other in neighboring planes).
Thus, this measurement presents a clear confirmation of the
theory involving ${\cal H}_4$.

Before continuing with 2342, we now turn to 2122.
Unlike 2342 above $T_{\rm NII}$, where 
minimizing ${\cal H}_4$ causes the spins to
point along the CuI--CuI bond, 
in 2122 the spins are believed to point colinearly along [110], i. e.
at $45^o$ with the Cu-Cu bond \cite{DV90}.
This implies that $H_{\rm A}^{\rm in,K}$ from Eq. (\ref{HAin}) enters into
the in-plane gap with a {\it negative} sign, 
and that the actual gap must also have
positive contributions from the interplanar exchange anisotropies, which 
dominate ${\cal H}_4$.
Our analysis \cite{details} indeed yields
\begin{eqnarray}
\label{ineq}
\omega_{\rm in}^2 &=& 2H_{\rm E} (H_{\rm A}^{\rm in,d} - H_{\rm A}^{\rm in,K})
\nonumber \\
&=& 32J(A_{\rm dip}+A_{\rm pdip})-8JS H_{\rm A}^{\rm in,K} \
,
\end{eqnarray}
where we used Eq. (\ref{Hd}).
 We now show that this equation is
reasonably fulfilled. We take
$\delta J = 0.04$ meV from Ref. \onlinecite{entin},
and thereby get $H_{\rm A}^{\rm in,K}=2 \times 10^{-6}$ meV.  
We also take $A_{\rm dip} \approx 2.7 \times 10^{-6}$ meV \cite{ERR}.
The experimental
value $\omega_{\rm in} \approx 0.048$ meV then implies that 
$A_{\rm pdip} \approx - 2 \times 10^{-6}$ meV. Writing $A_{\rm pdip}=2S^2
\delta J_{\rm int}$, where $\delta J_{\rm int}$ is the anisotropy
of the interlayer exchange interaction\onlinecite{structure}, we
find $\delta J_{\rm int} \sim -10^{-5}$ meV.
Assuming that $|\delta J_{\rm int}| \sim 10^{-4} J_{\rm int}$,
we estimate an interplanar exchange energy $J_{\rm int} \sim $
0.1 meV, of the same order of magnitude as the estimate
for La$_2$CuO$_4$, $J_{\rm int} \approx$ 0.25 meV \cite{structure}.
Thus we have corroborated Eq. (\ref{ineq}).

We now return to the
low$-T$ phase ($T$$<$$T_{\rm NII}$) in 2342. This phase has six spins
per unit cell, and four
low energy modes\cite{kim}, of which two, denoted in Ref. \onlinecite{kim}
by $\omega_3$
and $\omega_4$, with energies above 5 meV, were
used there to measure the parameter $j_{\rm eff}$
related to the I--II biquadratic coupling.
The two new modes at lower frequency, which we call $\omega^<_{\rm in}$ and
$\omega^<_{\rm out}$, represent respectively in-plane and out-of-plane
fluctuations, mostly on the CuII ions (these were denoted $\omega_1$ and
$\omega_2$ in Ref. \cite{kim}).
$\omega^<_{\rm out}$
concerns out-of-plane
fluctuations of the CuII spins. These are practically not affected by 
$H_{\rm A}^{\rm in,K}$,
and $\omega^<_{\rm out}$ is equal to $\omega_2$
in Kim {\it et al.}\cite{kim}. Using the
parameters as listed in\cite{kim}, we predict
$\omega^<_{\rm out}=1.77$ meV, which
agrees nicely with the present result 422.5 GHz (=1.747 meV).

    The effective in-plane anisotropy energy was neglected in the
theoretical expressions in Ref. \onlinecite{kim},
because it had only insignificant
effects on the modes studied there.  In contrast, 
$\omega^<_{\rm in}$ is determined by this energy.
Accordingly we use Eq. (\ref{gap}), but now we have
$H_{\rm E}=4SJ_{\rm II}$, where $J_{\rm II}\approx 10.5$ meV
is the CuII-CuII exchange energy \cite{chou}, and $H_{\rm A}^{\rm in}$
has contributions from {\it both} CuI and CuII.
Our new spin wave analysis yields  \cite{details}
\begin{equation}
H_{\rm A}^{\rm in}=(8k_{\rm eff}/S) J/(J-J_{\rm I-II} + 2J_{\rm II}) \ ,
\label{HAin1}
\end{equation}
where $J_{\rm I-II}=-10$ meV \cite{tornow} and 
the relevant anisotropies are contained
in the parameter $k_{\rm eff}$, where 
\begin{equation}
k_{\rm eff}=k+ \case 1/2 A-K_{\rm II}.
\label{keff}
\end{equation}
The first term, $k$, contains the four-fold anisotropy energy of the CuI's,
and contributions from the ferromagnetic canting of 
both the CuI and CuII. Unlike the higher $T$ phase, the canting of the CuII
is now not negligible, and we recover
\cite{kastner}
\begin{eqnarray}
k=2K+8 J_{pd}^2 M_{\rm I}^{\dagger 2} [0.53/(8J_{\rm II})]
\label{kk}
\end{eqnarray}
($M_{\rm I}^{\dagger}$ 
is the staggered moment on
the CuI's, and $J_{pd}$ is the pseudodipolar part of the CuI--CuII
exchange).
The last two terms in Eq. (\ref{keff}) come from the CuII spins.
Since the spin structure of the CuII ions is similar 
to that of the Cu's in 2122
(the spins point at 45$^o$ to the CuII-CuII bond), these two terms are
analogous to those in Eq. (\ref{ineq}). Using Eq. (\ref{HAin}), with
$|J_{\parallel}-J_{\perp}| \sim 10^{-4}J$ for the CuII's, we estimate
that $K_{\rm II} \sim K/10$.
At low $T$, $k_{\rm eff}$ is thus dominated by the
interplanar dipolar term $\case 1/2 A$.
Assuming for simplicity only real dipolar interactions, we have
$A=3(g \mu_B M_{\rm II}^{\dagger})^2 X$, 
where $X$ is the lattice sum in Eq. (10) of Ref. \onlinecite{structure},
which we evaluated as $7 \times 10^{-4} \AA^{-3}$. Thus, at $T=0$ we estimate
$k_{\rm eff} \approx 24 \times 10^{-6}$ meV, in reasonable
agreement with the static value  
\cite{kastner}.
The mysterious dramatic increase in $k_{\rm eff}$ observed in Ref.
\onlinecite{kastner} below
$T_{\rm NII}$ is thus explained by the additional dominant term $\case 1/2 A$
(which is proportional to $M_{\rm II}^{\dagger 2}$).
Using this estimate in Eq. (\ref{HAin1})
yields $\omega^<_{\rm in} \approx 0.12$
meV, not far from the experimental value 0.15 meV.



The g-tensor is calculated from ${\cal H}=\mu_B {\bf H} \cdot ({\bf L} + 2 {
\bf S})$.  The quantum average is calculated for the ground state, so one
has $g=2+g_L$, where $<{\bf L}>=g_L<{\bf S}>$. The latter is calculated
perturbatively, with the spin-orbit term $\lambda {\bf L}\cdot{\bf S}$, i.
e.
\begin{eqnarray}
<L_\alpha>&=&<L_\alpha (1/E) \lambda {\bf L}\cdot{\bf S}> +
<\lambda {\bf L}\cdot{\bf S} (1/E) L_\alpha>
\nonumber \\
&=& 2 \lambda <S_\alpha><L_\alpha (1/E) L_\alpha>,
\end{eqnarray}
where $E$ represents the energy of an intermediate state.

Thus,
\begin{eqnarray}
g_{x,y} &=& 2 + 2 \lambda/e_{x,y} \nonumber
\\
g_z &=& 2 + 8 \lambda/e_z.
\end{eqnarray}
Taking $e_x=e_y=e_z=1.8$ eV and $\lambda = 0.1$ eV gives $g_{x,y}=2.11$,
$g_z=2.45$, in reasonable agreement with  the present results 2.08 and 2.30.

In conclusion, we have measured the low energy excitations of the $S$=$\frac
{1}{2}$ tetragonal antiferromagnets, Sr$_{2}$CuO$_{2}$Cl$_{2}$ and
Sr$_{2}$Cu$_{3}$O$_{4}$Cl$_{2}$ using an ESR spectrometer covering a wide
range of frequency and magnetic field. At 70 K for 2342,
we have been successful in observing
the quantum in-plane energy gap at $H$=0 predicted theoretically.
The other in-plane gaps which we have measured
reflect additional anisotropies, which we have shown to 
have the expected orders of magnitude.

This work was supported by the MR Science Research Program of RIKEN, a
Grant-in-Aid for Scientific Research from the Japanese Ministry of
Education, Science, Sports and Culture, the U.S.-Israel Binational Science
Foundation (at TAU, MIT and Penn.), 
and the MRSEC Program of the National Science Foundation
under Grant No. DMR 9808941(at MIT).  Z. H. is supported by the
Research Fellowships of the Japan Society for the Promotion of
Science for Young Scientists.  J.S. is supported by the
Special Researcher's Basic Science Program of RIKEN.

\begin{table}[h]
\centering
\caption{Experimental results.}
\begin{minipage}{8.5 cm}
\renewcommand{\arraystretch}{1.2}
\begin{tabular}{c c c c c c}
&$T$, K & ${\bf H} \parallel$ & $g$ & $\nu(0)$, GHz & $\omega$, meV  \\
\hline
2342 &70  & [110]&$2.083(3)$&16.0(9)&$\omega_{\rm in}=0.066(4)$ \\ 
\hline
2342 & 1.5  & [110]&$2.080(2)$&36.1(6)&$\omega_{\rm in}^<=0.149(3)$ \\
2342 & 1.5 & [001]&$2.301(1)$& 422.5(1)&$\omega_{\rm out}^<=1.7473(4)$ \\
\hline
2122 & 5 & [110]&$2.046(2)$&11.6(7)& $\omega_{\rm in}=0.048(3)$ \\
\end{tabular}
\end{minipage}
\label{TableI}
\end{table}

\begin{figure}
\caption{The electron spin resonance signal observed in a single crystal of
Sr$_{2}$Cu$_{3}$O$_{4}$Cl$_{2}$ at 70 K and at several frequencies. 
${\bf H} \parallel$ [110].}
\label{fig1}
\end{figure}

\begin{figure}
\caption{The frequency versus magnetic field plots of the ESR signals
observed in a single crystal of Sr$_{2}$Cu$_{3}$O$_{4}$Cl$_{2}$ at 1.5 K for
the two field directions, and at 70 K for ${\bf H} \parallel$ [110].
Inset: same data, $\nu^2$ versus $H^2$.}
\label{fig2}
\end{figure}


\begin{figure}
\caption{The frequency dependence of the resonance field in a single crystal
of Sr$_{2}$CuO$_{2}$Cl$_{2}$ obtained at 5 K for ${\bf H} \parallel$ [110].
The dotted line is drawn through the origin, for comparison.
Inset: same data, $\nu^2$ versus $H^2$.}
\label{fig4}
\end{figure}


\begin{references}
\bibitem{renorm} As discused in Ref. \onlinecite{kim}, theoretical expansions
in powers of $1/S$ usually imply some quantum renormalization factors for
$\omega$. Since Eq. (\ref{HAin}) already represents a quantum correction,
we have not included these factors. However, this might imply an
uncertainty in the theoretical values, which could involve a factor of up
to 2.
\bibitem{ERR}
In Ref. \onlinecite{structure} the constant $A_{\rm dip}$ was
evaluated using $S=0.2$ as mentioned in their Ref. 18. However, we
prefer to use $S=0.3$.  Since $A_{\rm dip}$ is proportional to
$S^2$, we set $A_{\rm dip}=1.2 \times 10^{-6} (0.3/0.2)^2$ meV.
\end{references}
\end{document}